\def\a{\alpha}

\def\sm{Standard Model\hspace {.05in}}
\def\wpwp{$W^+W^+\hbox{ }$}
\def\wqwq{$W^+W^+ + W^-W^- \hbox{ }$}
\def\wpmz{$W^{\pm}Z \hbox{ }$}
\def\lag5{\mbox{${\cal L}_5\hbox{ }$}}

\def\hm{Higgs mechanism\hspace {.05in}}
\def\u1{$U(1)\hbox{ }$}

\def\su2xu1{$SU(2)_L \times U(1)_Y\hbox{ }$}
\def\rpp{$\rho \pi \pi\hbox{ }$}
\def\rhotc{$\rho_{TC}\hbox{ }$}
\def\ro{$\rho\hbox{ }$}
\def\qrho{``$\rho$''\hbox{ }}
\def\invfb{fb$^{-1}$\hspace{.05in}}
\def\qqb{$\overline{q}q \hbox{ }$}
\def\ttb{$\overline{t}t \hbox{ }$}
\def\ttt{$10^{33}$ cm$^{-2}$ sec$^{-1}\hbox{ }$}
\def\tttt{$10^{34}$ cm$^{-2}$ sec$^{-1}\hbox{ }$}

\def\wlwl{$W_LW_L\hbox{ }$}

\def\m5{$M_5\hbox{ }$}

\def\wpmz{$W^{\pm}Z\hbox{ }$}
\def\alphaw{$\alpha_W\hbox{ }$}
\def\alphaw2{$\alpha_W^2\hbox{ }$}

\def\oalphaw2{$O(\alpha_W^2)\hbox{ }$}

\def\pp{$\pi \pi\hbox{ }$}
\def\wpwmz{$w^+,w^-,z\hbox{ }$}
\def\wpwmzl{$W^+_L,W^-_L,Z_L\hbox{ }$}
\def\a11{$a_{11}\hbox{ }$}
\def\a20{$a_{20}\hbox{ }$}

\documentstyle[ichep,12pt]{article}
\title{\normalsize
August, 1992   \hfill    LBL-32938 \\
\vskip .2in
COMPLEMENTARITY OF RESONANT AND NONRESONANT STRONG WW
SCATTERING AT SSC AND LHC
\thanks
{This work was supported by the
Division of High
Energy Physics of the U.S. Department of Energy under Contract
DE-AC03-76SF00098.}
\\ (To be published in Proc. XXVI Intl. Conf. on High Energy Physics,\\
Dallas, Texas, August 1992)
 }
\author{ Michael S. Chanowitz\\
        Lawrence Berkeley Laboratory\\
        Berkeley CA 94720}
\begin{document}
\finalcopy

\maketitle

\abstract{
Signals and backgrounds for strong $WW$ scattering at the SSC and LHC are
considered. Complementarity of resonant signals in the $I=1$ $WZ$ channel and
nonresonant signals in the $I=2$ \wpwp channel is illustrated using a chiral
lagrangian with a $J=1$ \qrho resonance.
Results are presented for purely leptonic final states in the $W^{\pm}Z$,
$W^+W^+ + W^-W^-$, and $ZZ$ channels.
} 

\vskip-1pc
\onehead{INTRODUCTION}

High energy physics today is in an extraordinary situation. The Standard Model
(SM) is reliable but incomplete. For its completion it {\em predicts} 1) that a
fifth force exists, 2) the mass range of the associated quanta, and 3) neither
the precise mass nor the interaction strength but the relation between them.
These properties are sufficient to guide the search. Like any prediction in
science, this one too may fail. If so we will make an equally important
discovery: a deeper theory hidden until now behind the SM, which will emerge by
the same experimental program that we will follow to find the fifth force if it
does exist. In this paper I assume the SM is correct. This presentation is
necessarily brief; a more complete review and bibliography will
appear elsewhere.$^1$

The Higgs mechanism is the feature of the SM that requires a fifth force and
implies its general properties. The \hm requires a new sector of
quanta with dynamics
\break
\newpage
\vglue2pt
\noindent
specified by an unknown Lagrangian I will call ${\cal L}_5$, that
spontaneously breaks $SU(2)_L \times U(1)_Y$,
giving rise to Goldstone bosons \wpwmz that
become the longitudinal gauge bosons \wpwmzl. By measuring \wlwl scattering at
$E\gg M_W$, we are effectively measuring $ww$ scattering
(i.e., the equivalence theorem) and are therefore probing the dynamics of
${\cal L}_5$.

Let $M_5$ be the typical mass scale of the quanta of ${\cal L}_5$.
Then the \wlwl scattering amplitudes
are determined by low energy theorems,$^{2,3}$ e.g., for the $J=0$ partial wave
$$
a_0(W_L^+W_L^-\to Z_LZ_L)={1\over\rho}{s\over 16\pi v^2}\eqno(1)
$$
(with $v=0.247$ TeV) in the energy domain
$$
M^2_W \ll s \ll \hbox{minimum}\{M^2_{5}, (4\pi v)^2\}\eqno(2)
$$
which may or may not exist in nature, depending on whether $M_{5}\gg M_W$.

Partial wave unitarity requires the linear growth of $|a_0|$ to be damped
before
it exceeds unity at a ``cutoff'' scale
$\Lambda_5 \leq 4\pi\sqrt{v} = 1.8$ TeV. The cutoff is enforced
by the \hm with $\Lambda_5 \simeq M_5$ where more precisely
$M_5$ is {\it the mass scale of the
quanta of \lag5 that make the \su2xu1 breaking condensate that engenders
$M_W$.}
If $M_5 \ll 1.8$ TeV then \lag5 is weak and its
quanta include one or more Higgs bosons with $M_5$ equal to the average Higgs
boson mass (weighted by contribution
to $v$). If $M_5 \geq 1$ TeV then \lag5 is strong, there
is strong $WW$ scattering for $s > 1$
TeV$^2$, and rather than Higgs bosons we expect a complex spectrum of quanta.
Resonance formation then occurs in attractive channels at the energy scale of
unitarity saturation, $a_J(M^2) \sim \hbox{O}(1)$, implying $M \sim 1$ - 3 TeV.

We detect a strong \lag5 by observing strong $WW$ resonances and/or
strong nonresonant $WW$ scattering. Fortunately the two
approaches are complementary: if the resonances are very heavy and difficult to
observe there will be large signals in nonresonant channels.

\onehead{COMPLEMENTARITY}

If \lag5 contains no light quanta $\ll 1$ TeV such as Higgs bosons or pseudo
Goldstone bosons, then in the absence of strong $WW$ resonances the leading
partial wave amplitudes, $a_{IJ}=a_{00},a_{11},a_{20}$, will smoothly
saturate unitarity. Strong scattering cross sections are then estimated by
extrapolating the low energy theorems. (The index $I$ refers to the diagonal
$SU(2)_{L+R}$ subgroup that is necessarily$^3$
a good symmetry of the Goldstone boson sector at low energy because
$\rho \simeq 1$.)

Models illustrating the smooth approach to the unitarity limit include the
``linear'' model$^2$, the K-matrix unitarization model$^4$,
scaled \pp data in nonresonant
channels$^{2,4,5}$, and effective Lagrangians incorporating dimension 6
operators and/or one loop corrections$^{6}$. These models provide
large signals in {\it nonresonant} channels but are conservative
in that they apply when
more dramatic signals from light quanta or strong resonances are absent.

It is instructive to compare the linear model with \pp scattering data.$^7$
The model agrees well in the
$I,J=0,0$ channel, probably a fortuitous result of the
attractive dynamics in that channel. The model underestimates $|a_{11}|$
and overestimates $|a_{20}|$, {\it both} because of the $\rho(770)$:
s-channel \ro exchange enhances $|a_{11}|$ while
$t$- and $u$-channel exchanges suppress $|a_{20}|$, implying a complementary
relationship between the two channels.

The effects of \ro exchange can be studied using a chiral Lagrangian with
chiral invariant \rpp interaction.$^{8}$
Figure 1 shows that the model fits \pp data for $|a_{11}|$ and $|a_{20}|$ very
well.

\vskip 6in
\begin{small}
\noindent Figure 1.The \ro chiral Lagrangian model compared with \pp scattering
data for $|a_{11}|$ and $\delta_{20}$ (W. Kilgore).
\end{small}

\noindent We will
use the model to explore the effect of an analogous \qrho resonance on \wlwl
scattering.

Consider for instance minimal technicolor with one techniquark
doublet. (Nonminimal models have lighter resonances which are more
easily observed.) For $N_{TC}=4$,
large $N$ scaling implies $(m_{\rho},\Gamma_{\rho})
=(1.78,0.33)$ TeV, while the heaviest \rhotc, for $N_{TC}=2$, has
$(m_{\rho},\Gamma_{\rho})=(2.52,0.92)$ TeV. Though unlikely according to
popular
prejudice, strong $WW$ resonances could be even heavier.
To explore

\vskip 5.7in
\begin{small}
\noindent Figure 2.
$|a_{11}|$ and $\delta_{20}$ for the chiral invariant \ro exchange model
with $m_\rho = 1.78$ (dashes), $m_\rho = 2.52$ (long dashes) and
$m_\rho = 4.0$ (dot-dash).  The nonresonant $K$-LET model is indicated
by the solid line.
\end{small}

\noindent that possibility I also consider a \qrho of mass 4 TeV, with a width
of 0.98 TeV determined assuming a ``$\rho$''$ww$ coupling equal to
$f_{\rho\pi\pi}$ from hadronic physics.
To ensure elastic unitarity the real parts are computed with
$\Gamma_{\rho}=0$ and the K-matrix prescription is then used to
compute
the imaginary parts.$^{9}$ For resonance dominance this prescription is
equivalent to the usual broad-resonance Breit-Wigner prescription, in which
the term $m_{\rho}\Gamma_{\rho}$ in the B-W denominator
is replaced by $\sqrt{s}\Gamma_{\rho}(\sqrt{s})$.

Figure 2 displays $|a_{11}|$ and $|a_{20}|$ for the
three \qrho cases and for the nonresonant K-matrix unitarization of the low
energy theorem amplitudes (K-LET). The 4 TeV \qrho is nearly
indistinguishable from the nonresonant K-LET model below 3 TeV.
The complementarity of the two
channels is evident: the $\rho_{TC}(1.78)$ provides a spectacular
signal in $a_{11}$ but suppresses the signal in $a_{20}$, while the
``$\rho$''(4.0) provides a minimal signal in $a_{11}$ but allows a large
signal to emerge in $a_{20}$.

The sign of the interference between the LET amplitude and resonance
exchange contributions depends on the resonance quantum numbers, but it is
generally true that
the amplitude approaches a smooth unitarization of the LET (e.g., the K-LET)
as $M_5 \rightarrow
\infty$. This is the limit in which the ``conservative'' nonresonant models
apply. A heavy \qrho is a worst case example since \qrho exchange
interferes destructively with the $a_{20}$ threshold amplitude
so
that the limiting behavior is approached from below as the \qrho mass is
increased. Resonances that interfere constructively in the channel would
provide
bigger signals.

\onehead{SIGNALS}

In this section I will briefly review signals and backgrounds
at the SSC and LHC, in the \wpmz, \wqwq, and $ZZ$ final states. Signals are
computed using the ET-EWA approximation (i.e., the combined equivalence
theorem-effective W approximation) with HMRSB structure
functions evaluated at $Q^2=M_W^2$. Only
final states with both gauge bosons decaying leptonically are considered.
Except for the
central jet veto$^4$ (CJV) considered in the \wpwp channel, the cuts apply only
to leptonic variables.

My criterion for a significant signal is
$$
\sigma^{\uparrow} = S/\sqrt{B}\ge 5\eqno(3)
$$
$$
\sigma^{\downarrow} = S/\sqrt{S+B}\ge 3,\eqno(4)
$$
respectively the standard deviations for the background to fluctuate
up to a false signal or for the signal plus background to
fluctuate down to the level of the background alone. The criterion is corrected
below for the acceptance in each channel. In addition $S \ge B$ is required
because of the theoretical uncertainty in the backgrounds,
expected to be known to within $\le \pm 30 \%$ after ``calibration'' studies
at the SSC and LHC.

\twohead{\underline{\qrho $\rightarrow WZ$}}

Consider \qrho $\rightarrow WZ \rightarrow l\nu + \overline ll$
with $l = e,\mu$ ($BR = 0.014$).
Production mechanisms are $\overline qq$ annihilation$^{10}$
and $WZ$ fusion$^3$, the latter computed using the chiral Lagrangian with
contributions from $a_{11}$ and $a_{20}$. Elastic
unitarity is imposed with the K-matrix prescription described above. The
dominant
background (and the only one considered here) is $\overline qq \rightarrow WZ$.
A simple cut on the WZ invariant mass and the gauge boson rapidities ($y_{W,Z}
\le 1.5$) suffices to demonstrate the observability of the signal. (The $WZ$
mass is measurable only up to a twofold ambiguity; a more realistic and
effective procedure is to cut on the charged lepton transverse momenta.)

The acceptance estimate$^{11}$ is $0.85 \times 0.95 \simeq 0.8$ so the
significance criterion for the uncorrected cross sections is
$\sigma^{\uparrow} \ge 5.5$ and $\sigma^{\downarrow} \ge 3.3$. The results
are shown in figure 3 and table 1.

\begin{small}
\noindent Table 1. Yields of $\rho^\pm$ signal and background events
per 10 fb$^{-1}$ at the SSC and LHC.
Cuts are $|y_W| < 1.5$, $|y_Z| < 1.5$, and $M_{WZ}$ as
indicated.
\end{small}
\begin{center}
\begin{tabular}{c|c|c|ccc}
$\sqrt{s}$&$M_{\rho}$&$M_{WZ}$&S&B &$\sigma^{\uparrow},
\sigma^{\downarrow}$\cr
\hline\hline
40&1.78 &$>$1.0 &30&9.3  &10, 4.8\cr
 TeV&2.52&$> 1.2$& 15&5.3&6.3, 3.3\cr
& 4.0&$>1.0$&10&5.3&4.4, 2.6\cr
\hline
16&1.78&$>1.0$&5.5&3.2&3.0, 1.9\cr
TeV&2.52& $>1.2$&1.7&1.6&1.4, 0.9\cr
&4.0&$>1.6$&0.5&0.5&0.7, 0.5\cr
\hline\hline
\end{tabular}
\end{center}

\vskip 10pt
\noindent With 10 \invfb at the SSC
the $\rho_{TC}(1.78)$ signal far exceeds the criterion, the $\rho_{TC}(2.52)$
signal just meets it, and the ``$\rho$''(4.0) requires 17
fb$^{-1}$. To just meet the criterion at the LHC, 33, 160, and 570 \invfb are
needed for the three cases respectively.

\twohead {\underline{\wqwq}}

The \wpwp channel has the largest leptonic branching ratio,
$\simeq 0.05$ to $e$'s and/or $\mu$'s, and no \qqb annihilation background.
The signature is striking: two isolated, high $p_T$, like-sign leptons in an
event with no other significant activity (jet or lepton) in the central region.
The dominant backgrounds are

\vskip 3in
\begin{small}
\noindent Figure 3. $WZ$ cross section at SSC and LHC with $|y_{W,Z}|<1.5$ for
$\rho(1.78)$ (solid), $\rho(2.52)$ (dashes), and \qqb background (dot-dash).
\end{small}

\begin{small}
\noindent Table 2. Cumulative effect of cuts on linear model
signal and background for
$W^+W^+$ only at the SSC.
Entries are events per 10 fb$^{-1}$.
\end{small}
\begin{center}
\begin{tabular}{c|cc}
Cut&Signal&Bkgd.\cr
\hline\hline
$|y_l| <2$ &71&560\cr
\hline
$p_{Tl} >0.1$ TeV&44&49\cr
\hline
$\cos \varphi_{ll} < - 0.975$&32&9.1\cr
\hline
CJV& 27&2.4\cr
\hline\hline
\end{tabular}
\end{center}

\vskip 10pt
\noindent the $\hbox{O}(\alpha_W^2)$$^{12}$ and
$\hbox{O}(\alpha_W
\alpha_S)$$^{13}$ amplitudes for $qq \rightarrow qqWW$. The former is
essentially the
\wpwp pair cross section from \su2xu1 gauge interactions,
computed using the
standard model with a light Higgs boson, e.g., $m_H \le 0.1$ TeV.
Other backgrounds, from $W^+W^-$ with
lepton charge mismeasured and from \ttb production, require detector
simulation.
Studies presented in the SDC TDR$^{11}$ show that they can be controlled.

A powerful set of cuts that efficiently though indirectly exploits the
longitudinal polarization of the signal has emerged from the efforts of three
collaborations.$^{4,5,14}$. The most useful variables are the
lepton transverse
momentum $p_{Tl}$ and the azimuthal angle between the two leptons
$\phi_{ll}$$^{14}$.
The CJV$^4$ also effectively exploits the $W$
polarization; since the CJV signal efficiency may be affected by QCD
corrections
I present results with and without it. The truth probably lies closer to
the results with CJV, but the necessary calculations have not been done.
The successive effect of these cuts is illustrated in table 2. Even without the
CJV they reduce the background by $\simeq {\rm O}(10^2)$ while decreasing the
signal by little more than a factor 2.

Assuming 85\% detection efficiency for a single isolated lepton,$^{11}$
eqs. (3-4) applied to the uncorrected yields become
$\sigma^{\uparrow}>6$ and $\sigma^{\downarrow}>3.5$. Typical results for
the linear,
K-LET, and scaled $\pi\pi$ data models are shown in table 3. In addition
to $y_l < 2$ the cuts are $p_{Tl}>0.2$ TeV and cos$\phi_{ll}<-0.975$ for the
linear and K-LET models and $p_{Tl}>0.1$ TeV and cos$\phi_{ll}<-0.90$ for the
$\pi\pi$ model. The observability criterion is exceeded by a large margin
at the SSC in all cases but one ---
the $\pi\pi$ model without CJV for which the
criterion is just satisfied. At the LHC both the signals and signal:background
ratios
are less favorable, and about 70 \invfb would be needed just to meet the
minimum criterion for $\sigma^{\downarrow}$.

Results for the chiral invariant \ro exchange model
are given in table 4. The cuts optimize the
signal without CJV. For the SSC they are $p_{Tl}>0.1$ TeV and
cos$\phi_{ll}<-0.925$ for $\rho(1.78)$ and $\rho(2.52)$, and
$p_{Tl}>0.2$ TeV and cos$\phi_{ll}<-0.975$ for $\rho(4.0)$.
Each case meets the minimum criterion with 10 \invfb
except $\rho(1.78)$ without CJV which would
require 17 fb$^{-1}$ but is readily observable with a big signal in the
$WZ$ channel (table 1). As expected from figure 2. the SSC yields for
$\rho(4.0)$ (table 4) are within 5\% of the K-LET yields (table 3).
Comparing with the $WZ$ yields in table 1, we see that 10 \invfb suffices
to detect the signal for any value of $m_{\rho}$ in at least one of the two
(complementary) channels.

The LHC cuts in table 4 are $p_{Tl}>0.15$ TeV and cos$\phi_{ll}<-0.95$
for all three models. The $\rho(1.78)$ signal would require 160 \invfb just to
meet the minimum criterion, while the $\rho(4.0)$ signal would require 55
fb$^{-1}$. With $\simeq$ 100 \invfb the LHC could meet the
minimum criterion for each model in at least one of the $WZ$ or \wpwp
channels,$^1$
assuming the relevant measurements can really be carried out at
$10^{34}$cm$^{-1}$s$^{-1}$ (and with the
efficiencies assumed here). In addition to instrumentation issues, the
\ttb backgrounds that have been studied at
\ttt have yet to be simulated at $10^{34}$.

\twohead {\underline{$ZZ$}}

Very heavy Higgs bosons and strong scattering into the $ZZ$ final state are
best
detected

\onecolumn
\begin{center}
\begin{small}
\begin{quotation}
\noindent Table 3. Signal ($S$) and background ($B$) \wqwq events per
10~\invfb at SSC and LHC for the indicated models. Cuts are specified
in the text.
\end{quotation}
\vskip 10pt
\end{small}
\begin{tabular}{c|c|ccc|ccc}
$\sqrt{s}$ & Model & \multicolumn{3}{c|}{No CJV}&
                \multicolumn{3}{c}{CJV}\cr
TeV  & &S &B &$\sigma^\uparrow, \sigma^\downarrow$ &S &B &$\sigma^\uparrow,
                                                \sigma^\downarrow$\cr
\hline\hline
& Linear   &30 &3.5 &16, 5.2  &26 &0.8 &29, 5\cr
40& K        &23 &3.5 &12, 4.4  &20 &0.8 &23, 4.4\cr
  & $\pi\pi$ &33 &26  &6.5, 4.3 &27 &6.5 &11, 4.7\cr
\hline
& Linear    &2.5 &0.5 & 3.5, 1.4 &2.1   &0.09  &6.9, 1.4\cr
16  & K         &2.0 &0.5 &2.8, 1.3  & 1.7  & 0.09 & 5.5, 1.3\cr
  & $\pi\pi$  &5.0 &5.4 &2.2, 1.6  &3.9   &1.0   &3.9, 1.8\cr
\hline\hline
\end{tabular}
\end{center}

\vskip 20pt
\begin{center}
\begin{small}
\begin{quotation}
\noindent Table 4. Signal ($S$) and background ($B$) \wqwq events per
10~\invfb at SSC and LHC for the \ro exchange model. Cuts are specified in the
text.
\end{quotation}
\end{small}
\vskip 10pt
\begin{tabular}{c|c|ccc|ccc}
$\sqrt{s}$ & $M_{\rho}$ &\multicolumn{3}{c|}{No CJV}&
                \multicolumn{3}{c}{CJV}\cr
TeV &TeV &S &B &$\sigma^\uparrow, \sigma^\downarrow$ &S &B &$\sigma^\uparrow,
                                                \sigma^\downarrow$\cr
\hline\hline
& 1.78  &22 &23  &4.6, 3.3 &18  &5.7 &7.6, 3.7\cr
40  & 2.52  &31 &23  &6.4, 4.2 &25  &5.7 &11, 4.5\cr
  & 4.0   &22 &3.5 &11, 4.3  &20  &0.8 &21, 4.4\cr
\hline
& 1.78  &1.8 &1.5 & 1.5, 1.0 &1.4  &0.3  &2.8, 1.1\cr
16  & 2.52  &2.4 &1.5 & 2.0, 1.2 &1.9  &0.3  &3.7, 1.3\cr
  & 4.0   &3.3 &1.5 &2.7, 1.5  &2.6  &0.3  &5.1, 1.5\cr
\hline\hline
\end{tabular}
\end{center}

\vskip 20pt
\begin{center}
\begin{small}
\begin{quotation}
\noindent Table 5. Linear model signals and background $ZZ$ events per 10
fb$^{-1}$
at SSC and LHC for various values of $m_t$.
Cuts are $|y_l|<2$ and $p_{Tl} > 75$ GeV.
For the SSC $M_{TZ}>$ 700 GeV and for the LHC $M_{TZ}>$ 600 GeV.
\end{quotation}
\end{small}
\vskip 10pt
\begin{tabular}{c|c|cc|c|cc}
$\sqrt{s}$&$m_{t}$&\multicolumn{2}{c|}{Signal}& Bkgd & $\sigma^\uparrow$ &
$\sigma^\downarrow$\cr
TeV& GeV& $gg$ &$WW$& &&\cr
\hline\hline
&100& 4.1 & 17.3&29.4&4.0&3.0\cr
40&150&10.1&17.3&30.3&5.0&3.6\cr
&200&16.7 &17.3& 32.2&6.0&4.2\cr
\hline
&100&0.75 &1.83 & 8.98&0.9&0.8\cr
16&150&1.72 & 1.83& 9.11 &1.2&1.0\cr
&200&2.41 & 1.83 & 9.49&1.4&1.2\cr
\hline\hline
\end{tabular}
\end{center}

\twocolumn
\noindent in the ``neutrino'' mode, $ZZ \rightarrow l^+l^- + \overline{\nu}\nu$
with $l=e$ or $\mu$. The net branching ratio from the $ZZ$ initial
state is 0.025, 6 times larger than the $l^+l^- + l^+l^-$ final state. The
signature --- a high $p_T$ $Z$ boson recoiling against missing $p_T$ with no
other
significant jet activity in the central rapidity region --- is experimentally
clean. Backgrounds from $Z + jets$ and from mismeasurement of the missing
$E_T$ have
been carefully studied and found to be controllable at \ttt for the SDC.$^{11}$
For the 1 TeV \sm Higgs boson with $m_t=150$ GeV, a cut of $y_l < 2$,
$p_{Tl}>75$ GeV and transverse mass $M_T > 600$ GeV provides a $14\sigma$
signal
with 96 signal events and 44 background events for 10 \invfb at the SSC.

If \lag5 is strongly interacting and if a single symmetry breaking condensate
gives mass to both the weak gauge bosons and to the top quark, then the $ZZ$
signal has {\em two} components.$^{15}$ Just as
$WW$ fusion probes the mass scale of the quanta which generate the
condensate that gives mass to $W$ and $Z$, $gg$ fusion via a
\ttb loop probes the quanta which generate the $t$ quark mass. If only one
condensate does both jobs, the $gg$ fusion contribution
enhances the strong scattering signal in the $ZZ$ final state.
This generalizes the two familiar Higgs boson production
mechanisms, $gg \rightarrow H$ and $WW \rightarrow H$,
to dynamical symmetry breaking with strong \lag5.

Results$^{15}$ are given in table 5.
Backgrounds considered are \qqb annihilation, $gg$ fusion, and the ${\rm
O}(\alpha_W^2)$ amplitude for $qq \rightarrow qqZZ$, the latter two
computed in the \sm with a light ($\leq 100$ GeV) Higgs boson. The
efficiency correction is offset by the additional contribution from $ZZ
\rightarrow l^+l^- + l^+l^-$ that is not included in table 5,
so eqs. (3-4) apply directly. For $m_t \geq 150$
GeV there are significant signals at the SSC with 10 \invfb thanks to the
big enhancement from $gg$ fusion.

The LHC signals with 10 \invfb are not significant.
To enforce $S\geq B$ the $p_{Tl}$ cut must be raised to 200 GeV, and
350 \invfb are then required to satisfy eqs. (3-4).
E.g., for $m_t = 150$ GeV the LHC with 350 \invfb yields 28 signal
and 31 background events, virtually identical to the SSC values in table 5 for
10 $fb^{-1}$. In addition the $Z + jets$ background
requires study at such high luminosity.

With luminosity above 10$^{33}$ at the SSC it
becomes possible to probe for multiple
condensates. E.g., if $m_t$ is generated by a light Higgs boson while
$M_W$ is generated dynamically$^{1,15}$ then only $WW$
fusion contributes to the $ZZ$ signal. For $m_t = 150$ GeV and 50 \invfb
the signal exceeds eqs. (3-4)
($\sigma^{\uparrow} = 7$ and $\sigma^{\downarrow} = 6$) and
differs by 3$\sigma$ from the one condensate model. We do not satisfy $S>B$
since $S/B=0.6$, but that may suffice given the years of experience likely to
precede such measurements.

It is unlikely that this measurement could be done at the LHC. To satisfy
$\sigma^{\uparrow} \geq 5$ for the two condensate model with $S/B=0.6$ would
require more than 1000 \invfb at the LHC.$^1$

\onehead {CONCLUSION}

The fifth force predicted by the \sm must begin to emerge at $\leq 2$ TeV in
$WW$ scattering. If that prediction fails, the \sm will be supplanted by a
deeper
theory that will begin to emerge in the same energy region. With 10 \invfb the
SSC has capability for the full range of possible signals: strong $WW$
scattering above 1 TeV or new quanta from \lag5 below 1 TeV. The strong
scattering signals can occur in complementary resonant and/or nonresonant
channels.

The practicability of measurements with $\geq$ \tttt is beyond the scope of
this
paper. In addition to accelerator and detector hardware questions there are
backgrounds --- some mentioned above --- which have been studied for \ttt but
require study at $10^{34}$. It may take years of experience
to learn to do physics in the 10$^{34}$ environment. If 100 \invfb data samples
are eventually achieved and the relevant backgrounds are overcome, the LHC
could
meet the minimum observability criterion for the models discussed here in at
least
one of the \wpwp and $WZ$ channels, while $\simeq 350$ \invfb would be needed
in
the $ZZ$ channel. Luminosity $\geq 10^{34}$ at the SSC would enable
the detailed studies of \lag5 that will be needed after the
initial discovery whether \lag5 is weak or strong. That program could extend
productively for several decades into the next century.

\vskip 10pt
Acknowledgements: I wish to thank Bill Kilgore for helping me to
understand the \ro exchange model, for suggesting a sensible unitarization
method, and for preparing the data compilations.

\onehead{REFERENCES}

\begin{enumerate}  {\itemsep0pt \parsep0pt \frenchspacing}

\item A more complete presentation of these results may be found in
M.S. Chanowitz, LBL-32846, 1992 (to be published in {\it
Perspectives on Higgs Physics}, N.Y.: World Sci.)
\item M.S. Chanowitz and M.K. Gaillard, {\it Nucl. Phys.} B261, 379 (1985).
\item M.S. Chanowitz, M. Golden, and H.M. Georgi, {\it Phys. Rev.} D36, 1490
(1987); {\it Phys. Rev. Lett.} 57, 2344 (1986).
\item V. Barger et al., {\it Phys. Rev.} D42, 3052 (1990).
\item M. Berger and M.S. Chanowitz, {\it Phys. Lett.} 263B, 509 (1991).

\newpage
\item T. Appelquist and C. Bernard, {\it Phys, Rev.} D22, 200 (1980); A.
Longhitano, {\it Phys. Rev.} D22, 1166 (1980);
J.F. Donoghue and C. Ramirez, {\it Phys. Lett.} 234B, 361 (1990);
A.Dobado, M.J.Herrero, and J.Terron, {\it Z. Phys.} C50, 205 (1991);
S.Dawson and G.Valencia, {\it Nucl. Phys.} B352, 27 (1991).
\item See fit $a$ in figure 4 of
J. Donoghue, C. Ramirez, and G. Valencia, {\it Phys. Rev.} D38, 2195 (1988).
\item S. Weinberg, {\it Phys. Rev.} 166, 1568 (1968).
\item This prescription is due to W. Kilgore.
\item E.Eichten et al., {\it Rev. Mod. Phys.} 56, 579 (1984).
\item Solenoidal Detector Collaboration, E.L. Berger et al.,
{\it Technical Design Report}, SDC-92-201, 1992.
\item D. Dicus and R. Vega, {\it Nucl. Phys.} B329, 533 (1990).
\item M.S. Chanowitz and M. Golden, {\it Phys. Rev. Lett.} 61, 1053 (1985);
{\it E 63}, 466 (1989);
D. Dicus and R. Vega, {\it Phys. Lett.} 217B, 194 (1989).
\item D. Dicus, J. Gunion, and R. Vega, {\it Phys. Lett.} 258B, 475 (1991);
D. Dicus, J. Gunion, L. Orr, and R. Vega, UCD-91-10, 1991.
\item M. Berger and M.S. Chanowitz, {\it Phys. Rev. Lett.} 68, 757 (1992).

\end{enumerate}

\end{document}